\documentclass[11pt,twoside]{article}

\usepackage{asp2006}
\usepackage{epsf}
\usepackage{lscape}
\begin{document}
\title{Characterizing Barred Galaxies in the Abell 901/902 Supercluster 
from STAGES}   

\author{I. Marinova,\altaffilmark{1} 
S. Jogee,\altaffilmark{1} 
D. Bacon,\altaffilmark{2}
M. Balogh,\altaffilmark{3} 
M. Barden,\altaffilmark{4}
F.~D.~Barazza,\altaffilmark{5} 
E. F. Bell,\altaffilmark{6}
A. B\"ohm,\altaffilmark{7} 
J. A. R. Caldwell,\altaffilmark{1 }
M. E. Gray,\altaffilmark{8}
B.~H\"au\ss ler,\altaffilmark{8}  
C. Heymans,\altaffilmark{9,10}
K. Jahnke,\altaffilmark{6}
E. van Kampen,\altaffilmark{4} 
S. Koposov,\altaffilmark{6}
K. Lane,\altaffilmark{8}
D. H. McIntosh,\altaffilmark{11}
K. Meisenheimer,\altaffilmark{6} 
C. Y. Peng,\altaffilmark{12,13}
H.-W.~Rix,\altaffilmark{6}
S. F. S\'anchez,\altaffilmark{14}
A. Taylor,\altaffilmark{15}
L. Wisotzki,\altaffilmark{7} 
C. Wolf,\altaffilmark{16} and
X. Zheng\altaffilmark{17}}   

\affil{\altaffilmark{1} The University of Texas, Department of Astronomy, Austin and Fort Davis, Texas, USA \\
\altaffilmark{2} Institute of Cosmology and Gravitation, University of Portsmouth, Portsmouth, UK \\
\altaffilmark{3} Department of Physics and Astronomy, University Of Waterloo, Waterloo, Canada\\
\altaffilmark{4} Institute for Astro- and Particle Physics, University of Innsbruck, Innsbruck, Austria\\
\altaffilmark{5} EPFL, Sauverny, Switzerland\\
\altaffilmark{6} Max-Planck-Institut f\"{u}r Astronomie, Heidelberg, Germany\\
\altaffilmark{7} Astrophysikalisches Insitut Potsdam, Potsdam, Germany\\
\altaffilmark{8} School of Physics and Astronomy, The University of Nottingham, Nottingham, UK\\
\altaffilmark{9} Department of Physics and Astronomy, University of British Columbia, Vancouver, Canada\\
\altaffilmark{10} Institut d'Astrophysique de Paris, Paris, France\\
\altaffilmark{11} Department of Astronomy, University of Massachusetts, Amherst, MA, USA\\
\altaffilmark{12} NRC Herzberg Institute of Astrophysics, Victoria, Canada\\
\altaffilmark{13} Space Telescope Science Institute, Baltimore, MD, USA\\
\altaffilmark{14} Centro Hispano Aleman de Calar Alto, Almeria, Spain\\
\altaffilmark{15} The Scottish Universities Physics Alliance (SUPA), Institute for
Astronomy, University of Edinburgh, Edinburgh, UK\\
\altaffilmark{16} Department of Astrophysics, University of Oxford, Oxford, UK\\
\altaffilmark{17} Purple Mountain Observatory, National Astronomical Observatories,
Chinese Academy of Sciences, Nanjing, China}

\begin{abstract} 
In dense clusters, higher densities at early epochs as well as
physical processes, such as ram pressure stripping and tidal interactions
become important, and can have direct
consequences for the evolution of bars and their host disks. 
To study bars and disks as a function of environment, 
we are using the STAGES ACS $HST$ survey of
the Abell 901/902 supercluster ($z\sim$~0.165), 
along with  earlier field studies based the SDSS and 
the Ohio State University Bright Spiral Galaxy Survey (OSUBSGS).
We explore the limitations of traditional methods for characterizing the
bar fraction, and in particular highlight uncertainties in disk galaxy 
selection in cluster environments. We present an alternative approach
for exploring the proportion of bars, and investigate the properties of 
bars as a function of host galaxy color, S{\'e}rsic index, stellar
  mass, star formation rate (SFR), specific SFR, 
and morphology.
\end{abstract}

\vspace{-1 cm}
\section{Introduction}
The most important internal driver of disk galaxy evolution are
stellar bars, because they efficiently redistribute angular momentum
between the disk and dark matter halo (e.g., Combes \& Sanders 1981; Weinberg 1985;
Athanassoula 2002).  
To put bars in a cosmological context, we must determine what effects
environment has on bar and galaxy evolution. The bar fraction and properties in
clusters relative to that found in field galaxies depend on several
factors, such as the epoch of bar formation, the higher densities in
clusters at early times leading to earlier collapse of dark matter
halos, and the relative importance of processes such as 
ram pressure stripping, galaxy tidal interactions, mergers, 
and galaxy harassment. For example, tidal
interactions can induce a bar in a dynamically cold disk, but they may
also heat the disk, making it less unstable to bar formation. 
Previous studies have found opposing results for the bar fraction in
isolated galaxies and those that are perturbed or in clusters (van den Bergh
2002; Varela et al. 2004).
We use our large supercluster sample of galaxies ($\sim$2000 over $M_{\rm V}$ -15.5 to -24.0) 
from the STAGES survey of the Abell 901/902 supercluster
($z\sim$~0.165) to investigate the fraction and properties of bars and their
host disks in a dense environment.

\section{STAGES Data and Sample}

\vspace{0 cm}
To study galaxies in a dense cluster environment, we use 
the Space Telescope A901/902 Galaxy Evolution Survey (STAGES; Gray et
al., in preparation). The Abell 901/902  supercluster ($z\sim$~0.165, 
number of galaxies per unit area N=250\,Mpc$^{-2}$) 
consists of three clusters: A901a, A901b, and A902 with
an average core separation of $\sim$~1\,Mpc. The A901
clusters show irregular X-ray morphologies, 
suggesting that they are not yet relaxed (Ebeling et al. 1996). 
The STAGES survey includes high resolution $HST$ ACS  F606W images 
(PSF $\sim$~$0.1\arcsec$, corresponding to $\sim$~300\,pc at
$z\sim$~0.165\footnote{We assume 
a flat cosmology with $\Omega_M = 1 - \Omega_{\Lambda} = 0.3$
and $H_{\rm 0}$ =70\,km~s$^{-1}$~Mpc$^{-1}$.}
) of the Abell 901/902 supercluster, along with spectrophotometric 
redshifts  of accuracy  $\delta_{\rm z}$/(1 + $z$)~$\sim$~0.02 
down to  $R_{\rm Vega}$= 24 from  the COMBO-17 survey (Wolf et al 2004).
Multi-wavelength coverage is available for this field from $GALEX$,
$Spitzer$, and XMM-Newton. 
Dark matter maps for the supercluster have been constructed using
gravitational lensing by Gray et al. (2002) and Heymans et al. (2008, MNRAS submitted). 
The cluster sample contains 798 bright, $M_{\rm V} \le -18.0$, galaxies.

\section{Method for Identification and Characterization of Bars}
To identify and characterize the properties of bars, we employ the
widely used method of fitting ellipses to the galaxy isophotes out to
sky level with the iraf task 'ELLIPSE' (e.g., Friedli et al. 1996;
Jogee et al. 1999; Knapen et al. 2000; Marinova \& Jogee 2007). We
generate plots of the surface brightness (SB), ellipticity ($e$), and
position angle (PA) as a function of radius for each galaxy. We also
plot overlays of the fitted ellipses onto the galaxy image. We use
both the overlays and radial plots to identify bars.  A galaxy is
identified as barred if (a)~the $e$ profile rises to a global maximum
while the PA stays constant and (b)~after the global max, the $e$
drops and the PA changes characterizing the disk region.  To ensure
reliable morphological classification, we exclude all galaxies with
outer $e > 0.5$ (i $> 60^{o}$).  With our PSF of $\sim$~$0.1\arcsec$,
($\sim$~300\,pc at $z\sim$~0.165), we cannot reliably detect bars with
diameter smaller than $\sim$~1.8\,kpc. However, such small bars are
usually nuclear bars, whereas we focus only on primary bars, which
have diameters greater than 2\,kpc.  When working in the rest-frame
optical, bars heavily obscured by dust and SF will be missed, while
such bars can be detected in the rest-frame NIR (Eskridge et al. 2000,
Knapen et al. 2000, Marinova \& Jogee 2007). Furthermore, partially
obscured bars can be missed in ellipse-fits because dust and SF along
the bar can cause the PA to vary marginally more than the 10$^{o}$
allowed by the constant PA criterion.  In fact, studies of nearby
galaxies suggest that the bar fraction increases by a factor of
$\sim$~1.3 in the NIR, compared to the optical (Marinova \& Jogee
2007).  As we do not have rest-frame NIR images for Abell 901/902, we
resort to a second method of classifying bars: visual classification.
The visual classification method tends to capture partially obscured
bars somewhat better than ellipse fits, because visual classification
takes into account, not only the stellar light in the bar, but also
secondary signatures, such as the shape of dust lanes, the overall
morphology of the disk, and spiral arms.
\section{Preliminary Analysis}

All bar studies to date carried out in field samples define the bar
fraction $f_{\rm bar}$ as the ratio (number of barred {\it disks}/
total number of {\it disks}).  An accurate determination of $f_{\rm
bar}$ therefore hinges on an accurate way to identify disk
galaxies. In field samples, both locally and at intermediate
redshifts, two techniques are widely used to identify disks:
S{\'e}rsic cuts ($n<$2.5) based on single component fits, and
luminosity-color cuts to isolate blue cloud galaxies from the red
sequence.  These methods have limitations even in field samples: the
S{\'e}rsic cut can miss bright disks with prominent bulges (where
$n>$2.5) and the luminosity-color selection misses bright disks with
red colors (caused by old stellar populations or dust-reddening).

In dense cluster environments, where disk galaxies can be red,  and where 
the luminosity function is dominated by faint dwarf galaxies, it becomes even 
harder to identify disks via either method.  
We illustrate these uncertainties in disk selection in the 
supercluster as follows. 
The barred galaxies identified in $\S$~3, from ellipse fit 
and visual classification, are plotted on the
color-luminosity  (Figure~1a) and S{\'e}rsic -luminosity (Figure~1b) planes.
Because bars are disk signatures, we can use the strongly barred 
galaxies missed by the two methods as a lower limit on their failure 
to select disk galaxies.  
For bright galaxies, we find that  46\% (45/97) and 40\% (39/97) 
of disks with prominent, visually-identified bars are missed, 
respectively, by the blue cloud color-luminosity cut 
and S{\'e}rsic cut.

It is clear  that the uncertainties in disk selection will cause
a large and dominant error in the optical bar fraction $f_{\rm bar}$ 
in clusters.
We therefore adopt the following approach in the A901/902 supercluster
(1)~We define a new quantity  $P_{\rm bar}$  as the proportion of  
{\it all galaxies} (rather than disk galaxies), which are barred.  
Thus, $P_{\rm bar}$ is not as heavily affected as $f_{\rm bar}$
by the uncertainties in disk selection. 
(2)~We explore how   $P_{\rm bar}$  and bar properties vary 
as a function of galaxy properties, such as color (Figure 1a), stellar 
mass (Figure 1c,d) , SFR (Figure 1c), specific SFR (Figure 1d), and
 bulge-to-disk ratio.
(3)~When comparing the frequency of bars in clusters to that in the field, 
we can only use $f_{\rm bar}$, since no measurements of $P_{\rm bar}$ 
exist in the field. We estimate $f_{\rm bar}$  by selecting disks via 
 visual classification, rather than from  color-luminosity  or 
S{\'e}rsic  cuts. Disk are identified visually based on  features 
such as spiral arms and stellar bars, or the presence of a bulge+disk 
from the light distribution.

\acknowledgements 
I.M. and S.J. acknowledge support from NSF grant AST 06-07748, NASA
LTSA grant NAG5-13063, as well as HST G0-10395. 

\begin{figure}[!b]
\begin{center}
\plotone{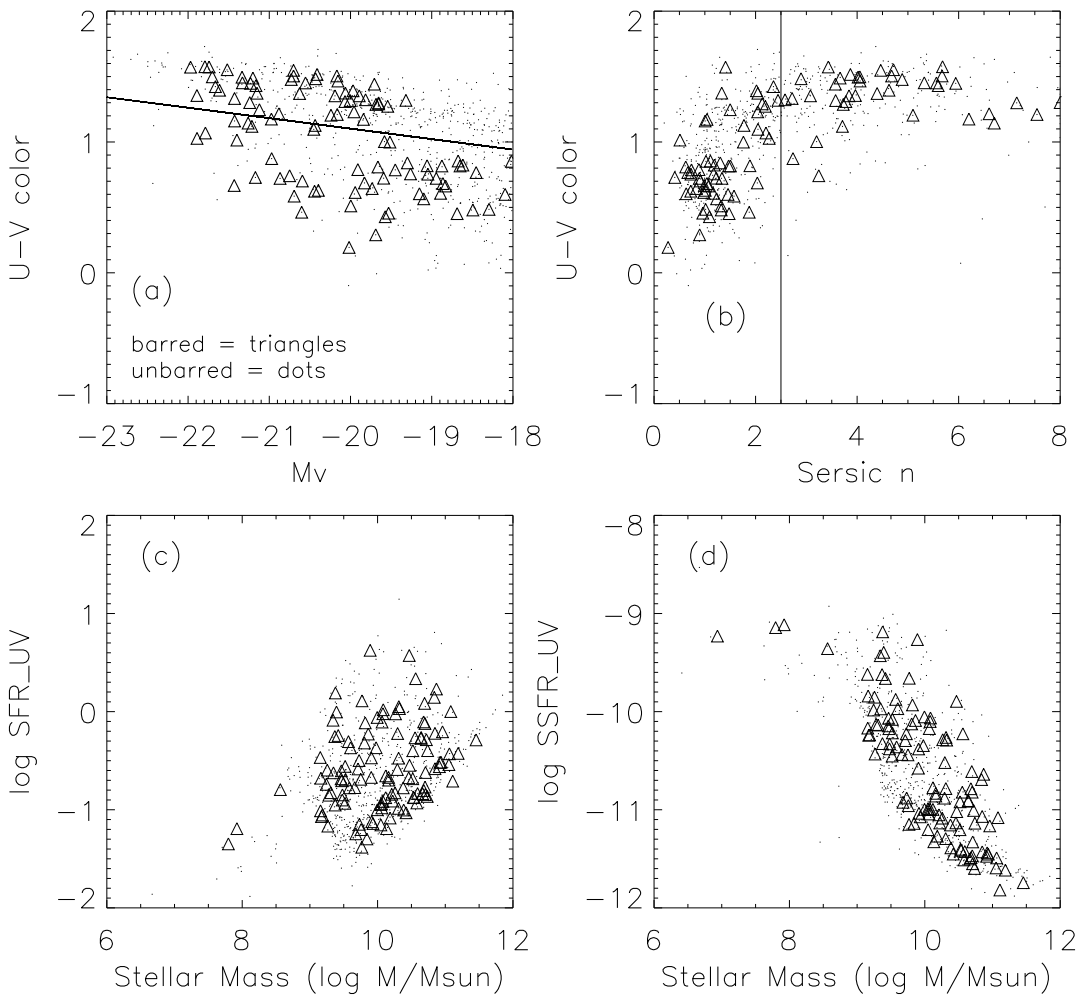}
\end{center}
\label{Figure 1:}
\vskip 0.2 in
\caption{(a)~Bright galaxies in rest-frame U-V vs. $M_{\rm V}$ plane. Galaxies with
 visually-identified strong bars are marked as triangles. 
 Many of these lie on the red sequence and
 would be missed by assuming disks lie only in the blue cloud.
(b)~Rest-frame U-V color vs. S{\'e}rsic $n$ plane. Many 
disks with prominent bars would be missed by a S{\'e}rsic cut
 $n<$2.5. (c)~SFR vs. stellar mass. (d)~Specific SFR  vs. stellar mass. Values for red sequence 
galaxies are upper limits.}
\end{figure}

\end{document}